\documentclass[11pt]{article}
\usepackage{amssymb}
\usepackage{pseudocode}

\newtheorem{theorem}{Theorem}[section]

\newtheorem{definition}[theorem]{Definition}

\newtheorem{claim}[theorem]{Claim}
\newtheorem{lemma}[theorem]{Lemma}

\newtheorem{corollary}[theorem]{Corollary}

\newcommand{\qedsymb}{\hfill{\rule{2mm}{2mm}}}
\newenvironment{proof}{\begin{trivlist}
\item[\hspace{\labelsep}{\sl\noindent Proof:\/}]
}{\qedsymb\end{trivlist}}

\def\balpha{\bar{\alpha}}
\def\bbeta{\bar{\beta}}

\def\Z{\mathbb{Z}}
\def\R{\mathbb{R}}
\def\mod{{\rm mod}}
\def\vol{{\rm vol}}
\def\half{{\frac{1}{2}}}

\newcommand\ontop[2]{{
\tiny\begin{array}{c} {#1} \\ {#2} \end{array} }}
\newcommand\ket[1]{{ |{#1} \rangle }}
\newcommand\C[1]{{ c_{\sf {#1}} }}
\newcommand\alternate[2]{{#1}}
\newcommand\omittedproof[1]{\alternate{#1}{Omitted.}}

\begin{document}

\title{\bf Quantum Computation and Lattice Problems }

\author{
Oded Regev
\thanks{Institute for Advanced Study, Princeton, NJ.
E-Mail: odedr@ias.edu. Research supported by NSF grant CCR-9987845.} }

\alternate{}{\date{}}
\maketitle

\begin{abstract}
We present the first explicit connection between quantum computation and lattice problems. Namely, we show a solution
to the Unique Shortest Vector Problem (SVP) under the assumption that there exists an algorithm that solves the hidden
subgroup problem on the dihedral group by coset sampling. Moreover, we solve the hidden subgroup problem on the
dihedral group by using an average case subset sum routine. By combining the two results, we get a quantum reduction
from $\Theta(n^{2.5})$-unique-SVP to the average case subset sum problem.
\end{abstract}

\section{Introduction}

Quantum computation is a computation model based on quantum physics. Assuming that the laws of nature as we know them
are true, this might allow us to build computers that are able to perform tasks that classical computers cannot perform
in any reasonable time. One task which quantum algorithms are known to perform much better than classical algorithm is
that of factoring large integers. The importance of this problem stems from its ubiquitous use in cryptographic
applications. While there are no known polynomial time classical algorithms for this problem, a groundbreaking result
of Shor from 1994~\cite{ShorFactor} showed a polynomial time quantum algorithm for factoring integers. In the same
paper, Shor showed an algorithm for finding the discrete log. However, despite enormous effort, we have only a few
other problems for which quantum algorithms provide an exponential speedup (e.g., \cite{HallgrenPell,HallgrenHCP}).
Other notable quantum algorithms such as Deutsch and Jozsa's algorithm~\cite{DeutschJozsa} and Simon's
algorithm~\cite{Simon} operate in the black box model. Grover's algorithm~\cite{Grover} provides a square root speedup
over classical algorithms.

The current search for new quantum algorithms concentrates on problems which are not known to be $NP$-hard. These
include the graph isomorphism problem and lattice problems. In this paper we are interested in lattice problems or
specifically, the unique shortest vector problem (SVP). A lattice is a set of all integral linear combinations of a set
of $n$ linearly independent vectors in $\R^n$. This set of $n$ vectors is known as a basis of the lattice. In the SVP
we are interested in finding the shortest nonzero vector in a lattice. In the $f(n)$-unique-SVP we are given the
additional promise that the shortest vector is shorter by a factor of at least $f(n)$ from all other non parallel
vectors. This problem also has important applications in cryptography. Namely, Ajtai and Dwork's
cryptosystem~\cite{AjtaiDwork} and the recent cryptosystem by Regev~\cite{RegevPKE} are based on the hardness of this
lattice problem.

A central problem in quantum computation is the hidden subgroup problem (HSP). Here, we are given a black box that
computes a function on elements of a group $G$. The function is known to be constant and distinct on left cosets of a
subgroup $H\leqslant G$ and our goal is to find $H$. Interestingly, almost all known quantum algorithms which run
super-polynomially faster than classical algorithms solve special cases of the HSP on Abelian groups. Also, it is known
that solving the HSP on the symmetric group leads to a solution to graph isomorphism~\cite{KoblerGraphIsomorphism}.
This motivated research into possible extensions of the HSP to noncommutative groups (see, e.g.,
\cite{GrigniSchulman01,HallgrenTashma00,RottelerWreathGroup,SanthaHiddenTranslation}). However, prior to this paper the
HSP on groups other than the symmetric group and Abelian groups had no known applications.

In this paper we will be interested in the HSP on the dihedral group. The dihedral group of order $2N$, denoted $D_N$,
is the group of symmetries of an $N$-sided regular polygon. It is isomorphic to the abstract group generated by the
element $\rho$ of order $n$ and the element $\tau$ of order 2 subject to the relation $\rho\tau = \tau \rho^{-1}$.
Although the dihedral group has a much simpler structure than the symmetric group, no efficient solution to the HSP on
the dihedral group is known. Ettinger and H{\o}yer~\cite{EttingerHoyerDihedral} showed that one can obtain sufficient
statistical {\em information} about the hidden subgroup with only a polynomial number of queries. However, there is no
efficient algorithm that solves the HSP using this information. Currently, the best known algorithm is due to Kuperberg
\cite{Kuperberg} and runs in subexponential time $2^{O(\sqrt{\log{N}})}$.

The following is the main theorem of this paper. The dihedral coset problem is described in the following paragraph.
\begin{theorem}\label{theorem_svp}
If there exists a solution to the dihedral coset problem with failure parameter ${\sf f}$ then there exists a quantum
algorithm that solves the $\Theta(n^{\frac{1}{2}+2{\sf f}})$-unique-SVP.
\end{theorem}
The input to the dihedral coset problem (DCP) is a tensor product of a polynomial number of registers. Each register is
in the state $\ket{0,x}+\ket{1,(x+d)~\mod~N}$ for some arbitrary $x\in\{0,\ldots,N-1\}$ and $d$ is the same for all
registers. These can also be thought of as cosets of the subgroup $\{(0,0),(1,d)\}$ in $D_N$. Our goal is to find the
value $d$. In addition, we say that the DCP has a failure parameter ${\sf f}$ if each of the registers with probability
at most $\frac{1}{(\log N)^{\sf f}}$ is in the state $\ket{b,x}$ for arbitrary $b,x$ instead of a coset state. We note
that any algorithm that solves the dihedral HSP by sampling cosets also solves the DCP for some failure parameter ${\sf
f}$. The reason is that since the algorithm samples only a polynomial number of cosets, we can take ${\sf f}$ to be
large enough such that with high probability all the registers are coset states. This is summarized in the following
corollary.
\begin{corollary}
If there exists a solution to the dihedral HSP that samples cosets (e.g., any solution using the `standard method')
then there exists a quantum algorithm that solves $poly(n)$-unique-SVP.
\end{corollary}

The following is the second main theorem of this paper. In the subset sum problem we are given two integers $t,N$ and a
set of numbers. We are asked to find a subset of the numbers that sums to $t$ modulo $N$. A legal input is an input for
which such a subset exists (a formal definition appears in Section~\ref{section_two_point}) and we are interested in
algorithms that solve a non-negligible part of the inputs:
\begin{theorem}\label{theorem_two_point}
If there exists an algorithm $S$ that solves $\frac{1}{poly(\log N)}$ of the legal subset sum inputs with parameter $N$
then there exists a solution to the DCP with failure parameter ${\sf f}=1$.
\end{theorem}
As shown in~\cite{EttingerHoyerDihedral}, the dihedral HSP can be reduced to the case where the subgroup is of the form
$\{(0,0),(1,d)\}$. Then, by sampling cosets we obtain states of the form $\ket{0,x}+\ket{1,(x+d)~\mod~N}$ with no
error. Hence,
\begin{corollary}
If there exists an algorithm $S$ that solves $\frac{1}{poly(\log N)}$ of the legal subset sum inputs with parameter $N$
then there exists a solution to the dihedral HSP.
\end{corollary}

Finally, the following is an immediate corollary of the two previous theorems:
\begin{corollary}
If there exists an algorithm that solves $\frac{1}{poly(\log N)}$ of the legal subset sum inputs with parameter $N$
then there exists a quantum algorithm for the $\Theta(n^{2.5})$-unique-SVP.
\end{corollary}
This result is known as a worst case to average case quantum reduction. Such reductions are already known in the
classical case \cite{Ajtai96atow,Cai98atow,CaiNerurkar97atow,Micciancio02atow,RegevPKE}. The exponent $2.5$ in our
reduction is better than the one in \cite{Ajtai96atow,Cai98atow,CaiNerurkar97atow,Micciancio02atow}. However, the
reduction in \cite{RegevPKE}, which appeared after the original publication of the current paper, further improves the
exponent to $1.5$ and hence subsumes our reduction. In addition, unlike the classical reductions, our subset sum
problems have a density of one, i.e., the size of the input set is very close to $\log N$. Therefore, some
cryptographic applications such as the one by Impagliazzo and Naor~\cite{ImpagliazzoNaor} cannot be used.

\subsection*{Intuitive overview}

Before proceeding to the main part of the paper, we describe our methods in a somewhat intuitive way. First, let us
describe the methods used in solving the unique-SVP. Recall that our solution is based on a solution to the DCP. We
begin by showing how such a solution can be used to solve a slightly different problem which we call the two point
problem. Instead of a superposition of two numbers with a fixed difference, our input consists of registers in a
superposition of two $n$-dimensional vectors with a fixed difference. Then, the idea is to create an input to the two
point problem in the following way. Start by creating a superposition of many lattice points and collapse the state to
just two lattice points whose difference is the shortest vector. Repeating this procedure creates an input to the two
point problem whose solution is the shortest vector.

Collapsing the state is performed by partitioning the space into cubes. Assume the partition has the property that in
each cube there are exactly two lattice points whose difference is the shortest vector. Then, we compute the cube in
which each point is located and measure the result. The state collapses to a superposition of just the two points
inside the cube we measured. The important thing is to make sure that exactly two points are located in each cube.
First, in order to make sure that the cubes are not aligned with the lattice, we randomly translate them. The length of
the cubes is proportional to the length of the shortest vector. Although the exact length of the shortest vector is
unknown, we can try several estimates until we find the right value. Since the lattice has a unique shortest vector,
all other nonparallel vectors are considerably longer and do not fit inside a cube. Therefore we know that the
difference between any two points inside the same cube is a multiple of the shortest vector. Still, this is not good
enough since instead of two points inside each box we are likely to have more points aligned along the shortest vector.
Hence, we space out the lattice: instead of creating a superposition of all the lattice points we create a
superposition of a subset of the points. The set of points created by this technique has the property that along the
direction of the shortest vector there are pairs of points whose difference is the shortest vector and the distance
between two such pairs is much larger than the shortest vector. As before, this can be done without knowing the
shortest vector by trying several possibilities.

The second part of the paper describes a solution to the DCP with failure parameter 1 which uses a solution to the
average case subset sum problem. Recall that we are given registers of the form $\ket{0,x}+\ket{1,(x+d)~\mod~N}$ where
$x\in \{0,\ldots,N-1\}$ is arbitrary and we wish to find $d\in \{0,\ldots,N-1\}$. Consider one such register. We begin
by applying the Fourier transform to the second part of the register (the one holding $x$ and $x+d$) and then measuring
it. If $a$ is the value we measured, the state collapses to a combination of the basis states $\ket{0}$ and $\ket{1}$
such that their phase difference is $2\pi \frac{ad}{N}$. If we were lucky enough to measure $a=1$, then the phase
difference is $2\pi \frac{d}{N}$ and by measuring this phase difference we can obtain an estimation on $d$. This,
however, happens with exponentially small probability. Since the phase is modulo $2\pi$, extracting the value $d$ is
much harder when $a$ is larger. Instead, we perform the same process on $r$ registers and let $a_1,\ldots,a_r$ be the
values we measure. The resulting tensor state includes a combination of all $2^r$ different $0,1$ sequences. The phase
of each sequence can be described as follows. By ignoring a fixed phase, we can assume that the phase of the sequence
$00\ldots 0$ is $0$. Then, the phase of the sequence $100\ldots 0$ is $2\pi \frac{a_1 d}{N}$ and in general, the phase
of the sequence $\alpha_1\alpha_2\ldots\alpha_r$ is $2\pi \frac{d}{N}$ multiplied by the sum of the values $a_i$ for
which $\alpha_i=1$. This indicates that we should try to measure the phase difference of two sequences whose sums
differ by $1$. However, although we can estimate the phase difference of one qubit, estimating the phase difference of
two arbitrary sequences is not possible.

We proceed by choosing $r$ to be very close to $\log N$. This creates a situation in which for almost every
$t\in\{0,\ldots,N-1\}$ there is a subset whose sum modulo $N$ is $t$ and in addition, there are not too many subsets
that sum to the same $t$ modulo $N$. Assume for simplicity that every $t$ has exactly one subset that sums to $t$
modulo $N$. We calculate for each sequence the value $\lfloor \frac{t}{2} \rfloor$ where $t$ is its sum. After
measuring the result, say $s$, we know that the state is a superposition of two sequences: one that sums to $2s$ and
one that sums to $2s+1$. Notice that since $a_1,\ldots,a_r$ are uniformly chosen between $\{0,\ldots,N-1\}$ we can use
them as an input to the subset sum algorithm. The key observation here is that the subset sum algorithm provides the
reverse mapping, i.e., from a value $t$ to a subset that sums to $t$. So, from $s$ we can find the sequence $\alpha_1$
that sums to $2s$ and the sequence $\alpha_2$ that sums to $2s+1$. Since we know that the state is a superposition of
$\ket{\alpha_1}$ and $\ket{\alpha_2}$ we can use a unitary transformation that transforms $\ket{\alpha_1}$ to $\ket{0}$
and $\ket{\alpha_2}$ to $\ket{1}$. Now, since the two states differ in one qubit, we can easily measure the phase
difference and obtain an estimate on $d$. This almost completes the description of the DCP algorithm. The estimate on
$d$ is only polynomially accurate but in order to find $d$ we need exponential accuracy. Hence, we repeat the same
process with pairs whose difference is higher. So, instead of choosing pairs of difference $1$ we choose pairs of
difference $2$ to get an estimate on $2d$, then $4$ to get an estimate on $4d$ and so on\footnote{This description is
very similar to the method of exponentially accurate phase estimation used in Kitaev's
algorithm~\cite{KitaevAbelianSubgroup}. Actually, our case is slightly more difficult because we cannot measure all the
multiples $2^i$. Nevertheless, we can measure enough multiples of the phase to guarantee exponential accuracy.}.

\subsection*{Outline}

The next section contains some notations that are used in this paper. The two main sections of this paper are
independent. In Section~\ref{section_svp} we prove Theorem~\ref{theorem_svp} and Section~\ref{section_two_point}
contains the proof of Theorem~\ref{theorem_two_point}.

\section{Preliminaries}

We denote the imaginary unit by $\imath$ and use the notation $e(x) = e^{2\pi \imath x}$. Occasionally, we omit the
normalization of quantum states. We use the term $n$-ball to refer to the $n$-dimensional solid body and the term
sphere to refer to its surface. We denote the set $\{1,\ldots,n\}$ by $[n]$. All logarithms are of base 2 unless
otherwise specified. We use $\delta_{ij}$ to denote the Kronecker delta, i.e., 1 if $i=j$ and 0 otherwise. A sequence
$\balpha\in \{0,1\}^r$ is identified with the set $\{i~|~\alpha_i=1\}$. Several constants appear in our proofs. To make
it easier to follow, we denote constants with a subscript that is somewhat related to their meaning. Specifically, in
Section~\ref{section_svp}, $\C{cub}$ is related to the cubes that partition the space, $\C{bal}$ is related to the
radius of the balls, and $\C{unq}$ appears in the guarantee of the unique shortest vector. Also, in
Section~\ref{section_two_point} we use $\C{r}$ in the definition of the parameter $r$, $\C{s}$ in our assumptions on
the subset sum subroutine and $\C{m}$ when we prove the existence of matchings.

The following is the formal definition of the DCP:
\begin{definition}
The input to the DCP with failure parameter ${\sf f}$ consists of $poly(\log N)$ registers. Each register is with
probability at least $1-\frac{1}{(\log N)^{\sf f}}$ in the state $\frac{1}{\sqrt{2}}(\ket{0,x}+\ket{1,(x+d)~\mod~N})$
on $1+\lceil \log N \rceil$ qubits where $x\in\{0,\ldots,N-1\}$ is arbitrary and $d$ is fixed. Otherwise, with
probability at most $\frac{1}{(\log N)^{\sf f}}$, its state is $\ket{b,x}$ where $b\in\{0,1\}$ and
$x\in\{0,\ldots,N-1\}$ are arbitrary. We call such a register a `bad' register. We say that an algorithm solves the DCP
if it outputs $d$ with probability $poly(\frac{1}{\log N})$ and time $poly(\log N)$.
\end{definition}

\section{A Quantum Algorithm for unique-SVP}
\label{section_svp}

In this section we prove Theorem~\ref{theorem_svp}. We begin by showing a simple reduction from the two point problem
to the DCP in Section \ref{sec_2pp}. We then prove a weaker version of Theorem~\ref{theorem_svp} with
$\Theta(n^{1+2{\sf f}})$ instead of $\Theta(n^{\frac{1}{2}+2{\sf f}})$ in Section \ref{sec_first_alg}. We complete the
proof of Theorem~\ref{theorem_svp} in Section \ref{sec_improved_alg}. Throughout this section, we use a failure
parameter ${\sf f}>0$ in order to make our results more general. The reader might find it easier to take ${\sf f}=1$.

\subsection{The Two Point Problem}\label{sec_2pp}

\begin{definition}
The input to the two point problem with failure parameter ${\sf f}$ consists of $poly(n\log M)$ registers. Each
register is with probability at least $1-\frac{1}{(n \log (2M))^{\sf f}}$ in the state
$\frac{1}{\sqrt{2}}(\ket{0,\bar{a}}+\ket{1,\bar{a}'})$ on $1+n \lceil \log M \rceil$ qubits where $\bar{a},\bar{a}'\in
\{0,\ldots,M-1\}^n$ are arbitrary such that $\bar{a}'-\bar{a}$ is fixed. Otherwise, with probability at most
$\frac{1}{(n \log (2M))^{\sf f}}$, its state is $\ket{b,\bar{a}}$ where $b\in\{0,1\}$ and $\bar{a} \in
\{0,\ldots,M-1\}^n$ are arbitrary. We say that an algorithm solves the two point problem if it outputs
$\bar{a}'-\bar{a}$ with probability $poly(\frac{1}{n\log M})$ and time $poly(n\log M)$.
\end{definition}

\begin{lemma}\label{n_dimensional_2pp}
If there exists an algorithm that solves the DCP with failure parameter $f$ then there is an algorithm that solves the
two point problem with failure parameter $f$.
\end{lemma}
\begin{proof}\omittedproof{
Consider the following mapping from $\{0,\ldots,M-1\}^n$ to $\{0,\ldots,(2M)^n-1\}$:
$$f(a_1,\ldots,a_n) = a_1+a_2\cdot 2M + \ldots + a_n (2M)^{n-1}.$$
Given an input to the two point problem, we create an input to the DCP by using the above mapping on the last $n\lceil
\log M \rceil$ qubits of each register. Hence, each register is with probability at least $1-\frac{1}{(n(\log 2M))^{\sf
f}}$ in state
 $$\frac{1}{\sqrt{2}} (\ket{0,f(\bar{a})} + \ket{1,f(\bar{a}')}).$$
The difference $f(\bar{a}')-f(\bar{a})$ is $(a'_1-a_1)+(a'_2-a_2)\cdot 2M + \ldots + (a'_n-a_n) (2M)^{n-1}$ and is
therefore fixed. Otherwise, with probability at most $\frac{1}{(n(\log 2M))^{\sf f}}$ the register is in the state
$\ket{b,f(\bar{a})}$ for arbitrary $b,\bar{a}$. This is a valid input to the DCP with $N=(2M)^{n}$ since the
probability of a bad register is at most $\frac{1}{(n \log (2M))^{\sf f}} = \frac{1}{(\log N)^{\sf f}}$.

Using the DCP algorithm with the above input we obtain the difference $b_1+b_2 \cdot 2M + \ldots + b_n (2M)^{n-1}$
where $b_i=a'_i-a_i$. In order to extract the $b_i$'s we add $M+M\cdot 2M + M(2M)^2+\ldots+M(2M)^{n-1}$. Extracting
$b_i$ from $(b_1+M)+(b_2+M) \cdot 2M + \ldots + (b_n+M) (2M)^{n-1}$ is possible since each $b_i + M$ is an integer in
the range $1$ to $2M-1$. The solution to the two point problem is the vector $(b_1,\ldots,b_n)$. }\end{proof}

\subsection{A Weaker Algorithm}\label{sec_first_alg}

We recall several facts about an LLL-reduced basis. Such a basis can be found for any lattice by using a polynomial
time algorithm~\cite{LLL}. Given a basis $\langle \bar{b}_1,\ldots,\bar{b}_n \rangle$, let $\langle
\bar{b}_1^*,\ldots,\bar{b}_n^* \rangle$ be its Gram-Schmidt orthogonalization. That is, $\bar{b}_i^*$ is the component
of $\bar{b}_i$ orthogonal to the subspace spanned by $\bar{b}_1,\ldots,\bar{b}_{i-1}$. An LLL reduced basis $\langle
\bar{b}_1,\ldots,\bar{b}_n \rangle$ satisfies that $\|\bar{b}_i^*\| \le \sqrt{2} \|\bar{b}_{i+1}^*\|$ and that for
$i>j$, $|\langle \bar{b}_i, \bar{b}_j^*\rangle| \le \frac{1}{2}\|\bar{b}_j^*\|^2 $. In addition, recall that $\min_i
\|\bar{b}_i^*\|$ is a lower bound on the length of the shortest vector. Since $\bar{b}_1^* = \bar{b}_1$ and
$\|\bar{b}_1^*\| \le 2^{(i-1)/2} \|\bar{b}_{i}^*\|$ we get that the vector $\bar{b}_1$ is at most $2^{(n-1)/2}$ times
longer than the shortest vector. Consider the representation of the LLL basis in the orthonormal basis $\langle
\frac{\bar{b}_1^*}{\|\bar{b}_1^*\|} , \ldots, \frac{\bar{b}_n^*}{\|\bar{b}_n^*\|} \rangle$. The vector $\bar{b}_i$ can
be written as $(b_{i1},b_{i2},\ldots,b_{ii},0,\ldots,0)$. Notice that $b_{ii}=\|\bar{b}_i^*\|$ and that $|b_{ij}| \le
\frac{1}{2} \|\bar{b}_j^*\|$ for every $i>j$. In the following, $\bar{u}$ denotes the shortest vector.

\begin{lemma}\label{svp_coordinates}
Consider the representation of the shortest vector $\bar{u}$ in the LLL-reduced lattice basis $\bar{u}=\sum_{i=1}^n u_i
\bar{b}_i$. Then, $|u_i| \le 2^{2n}$ for $i\in [n]$.
\end{lemma}
\begin{proof}\omittedproof{
Changing to the orthonormal basis, $\bar{u}=\sum_{i=1}^n u_i \bar{b}_i = \sum_{i=1}^n (\sum_{j=i}^{n} u_j
b_{j,i})\frac{\bar{b}_i^*}{\|\bar{b}_i^*\|}$. In addition, we know that $\|\bar{b}_i^*\| \ge 2^{-(i-1)/2}
\|\bar{b}_1^*\| \ge 2^{-n} \|\bar{u}\|$. Hence, $|\sum_{j=i}^{n} u_j b_{j,i}| \le 2^n \|\bar{b}_i^*\|$ for every $i\in
[n]$. By taking $i=n$ we get that $|u_n|$ is at most $2^n$. We continue inductively and show that $|u_k| \le 2^{2n-k}$.
Assume that the claim holds for $u_{k+1},\ldots,u_n$. Then, $|\sum_{j=k+1}^{n} u_j b_{j,k}| \le \frac{1}{2}
|\sum_{j=k+1}^n u_j| \|\bar{b}_k^*\| \le \frac{1}{2} (\sum_{j=k+1}^n 2^{2n-j}) \|\bar{b}_k^*\| \le \frac{1}{2} \cdot
2^{2n-k} \|\bar{b}_k^*\|$. By the triangle inequality, $|u_k b_{k,k}| \le |\sum_{j=k+1}^{n} u_j b_{j,k}| +
|\sum_{j=k}^{n} u_j b_{j,k}| \le (\frac{1}{2} 2^{2n-k} + 2^n) \|\bar{b}_k^*\| \le 2^{2n-k}\|\bar{b}_k^*\|$ and the
proof is completed.
}\end{proof}

Let $p > n^{2+2{\sf f}}$ be any fixed prime. The following is the main lemma of this section:

\begin{lemma}\label{main_lattice_lemma}
For any ${\sf f}>0$ let $\bar{u}=\sum_{i=1}^{n} u_i \bar{b}_i$ be the shortest lattice vector in a $(\C{unq} n^{1+2{\sf
f}})$-unique lattice where $\C{unq}>0$ is a constant. If there exists a solution to the two point problem with failure
parameter ${\sf f}$ then there exists a quantum algorithm that given this lattice and three integers $l,m,i_0$ returns
$(u_1,\ldots,u_{i_0-1},\frac{u_{i_0}-m}{p},u_{i_0+1},\ldots,u_n)$ with probability $1/poly(n)$ if the following
conditions hold: $\|\bar{u}\| \le l \le 2\|\bar{u}\|$, $u_{i_0}\equiv m~(\mod~p)$ and $1\le m\le p-1$.
\end{lemma}

We first show how this lemma implies Theorem~\ref{theorem_svp} with $\Theta(n^{1+2{\sf f}})$ by describing the SVP
algorithm. According to Lemma~\ref{n_dimensional_2pp} and the assumption of the theorem, there exists a solution to the
two point problem with failure parameter ${\sf f}$. Hence, Lemma~\ref{main_lattice_lemma} implies that there exists an
algorithm that given the right values of $l,m,i_0$ outputs
$(u_1,\ldots,u_{i_0-1},\frac{u_{i_0}-m}{p},u_{i_0+1},\ldots,u_n)$. The value $l$ is an estimate of the length of the
shortest vector $\bar{u}$. Because the LLL algorithm gives a $2^{(n-1)/2}$-approximation to the length of the shortest
vector, one of $(n-1)/2$ different values of $l$ is as required. In addition, since $\bar{u}$ is the shortest vector,
$\bar{u}/p$ cannot be a lattice vector and therefore there exists an $i_0$ such that $u_{i_0}\not\equiv 0~(\mod~p)$.
Hence, there are only $O(pn^2)$ possible values for $l,m$ and $i_0$. With each of these values the SVP algorithm calls
the algorithm of Lemma~\ref{main_lattice_lemma} a polynomial number of times. With high probability in one of these
calls the algorithm returns the vector $(u_1,\ldots,u_{i_0-1},\frac{u_{i_0}-m}{p},u_{i_0+1},\ldots,u_n)$ from which
$\bar{u}$ can be extracted. The results of the other calls can be easily discarded because they are either longer
lattice vectors or non-lattice vectors.

\begin{proof}(of Lemma~\ref{main_lattice_lemma})
We start by applying the LLL algorithm to the unique lattice in order to create a reduced basis. Denote the resulting
basis by $\langle \bar{b}_1,\ldots,\bar{b}_n \rangle$. Let $\langle \bar{e}_1,\ldots,\bar{e}_n \rangle$ be the standard
orthonormal basis of $\R^n$.

Let $w_1,\ldots,w_n$ be $n$ real values in $[0,1)$ and let $M=2^{4n}$. Assume without loss of generality that $i_0=1$.
The function $f$ is defined as $f(t,\bar{a})=(a_1 p + t m) \bar{b}_1 + \sum_{i=2}^n {a_i \bar{b}_i}$ where $t\in
\{0,1\}$ and $\bar{a}=(a_1,\ldots,a_n) \in {\cal A} = \{0,\ldots,M-1\}^n$. It maps the elements of $\{0,1\}\times {\cal
A}$ to lattice points. In addition, consider a lattice vector $\bar{v}$ represented in the orthonormal basis $\bar{v} =
\sum_{i=1}^n v_i \bar{e}_i$. The function $g$ maps $\bar{v}$ to the vector $(\lfloor v_1/(\C{cub} n^{\half+2{\sf f}}
\cdot l)-w_1 \rfloor , \ldots , \lfloor v_n/(\C{cub} n^{\half+2{\sf f}} \cdot l) - w_n \rfloor)$ in $\Z^n$ where the
constant $\C{cub}>0$ will be specified later.

In the following, we describe a routine that creates one register in the input to the two point problem that hides the
difference $(u_1,\ldots,u_{i_0-1},\frac{u_{i_0}-m}{p},u_{i_0+1},\ldots,u_n)$. We call the routine $poly(n\log
M)=poly(n)$ times in order to create a complete input to the two point problem. We then call the two point algorithm
and output its result. This completes the proof of the lemma since with probability $1/poly(n\log M)=1/poly(n)$ our
output is correct.

The routine starts by choosing $w_1,\ldots, w_n$ uniformly from $[0,1)$. We create the state
 $$ \frac{1}{\sqrt{2M^n}}\sum_{t\in \{0,1\}, \bar{a} \in {\cal A}} \ket{t,\bar{a}}. $$
Then, we compute the function $F=g\circ f$ and measure the result, say $r_1,\ldots,r_n$. The state collapses to
(normalization omitted)
 $$\sum_{\ontop{t\in\{0,1\}~~\bar{a} \in {\cal A}}{F(t,\bar{a}) = (r_1,\ldots,r_n) }} \ket{t,\bar{a}} \ket{r_1,\ldots,r_n}. $$
This completes the description of the routine. Its correctness is shown in the next two claims.

\begin{claim}
For every $\bar{r} \in \Z^n$, there is at most one element of the form $(0,\bar{a})$ and at most one element of the
form $(1,\bar{a}')$ that get mapped to $\bar{r}$ by $F$. Moreover, if both $(0,\bar{a})$ and $(1,\bar{a}')$ get mapped
to $\bar{r}$ then $\bar{a}'-\bar{a}$ is the vector $(\frac{u_1-m}{p},u_2,\ldots,u_m)$.
\end{claim}
\begin{proof}
Consider two different lattice points in the image of $f$, $\bar{v}=f(t,\bar{a})$ and $\bar{v}'=f(t',\bar{a}')$, that
get mapped to $\bar{r}$ by $g$. Let $\bar{v}=\sum_{i=1}^n v_i \bar{e}_i$ and $\bar{v}'=\sum_{i=1}^n v'_i \bar{e}_i$ be
their representation in the orthonormal basis. If $\bar{v}'-\bar{v}$ is not a multiple of the shortest vector, then
$\|\bar{v}'-\bar{v}\| > \C{unq} n^{1+2{\sf f}} \|\bar{u}\| \ge \frac{1}{2} \C{unq} n^{1+2{\sf f}} \cdot l$. Therefore,
there exists a coordinate $i\in [n]$ such that $|v'_i-v_i| \ge \frac{1}{2} \C{unq} n^{\half+2{\sf f}} \cdot l$ and for
$\C{unq} > 2\C{cub}$ this implies $g(\bar{v}) \neq g(\bar{v}')$ no matter how $w_1,\ldots,w_n$ are chosen. Hence,
$\bar{v}'-\bar{v} = k\cdot \bar{u}$ for some integer $k\neq 0$. By considering the first coordinate of
$\bar{v}'-\bar{v}$ in the lattice basis we get that $(a'_1 p + t'm)-(a_1 p + tm) \equiv k\cdot m~(\mod~p)$. This
implies that $k \equiv t'-t~(\mod~p)$. If $t=t'$ then $k\equiv 0~(\mod~p)$ which implies that $|k| \ge p$. Thus,
$\|\bar{v}'-\bar{v}\| \ge p\|\bar{u}\| > \C{cub} n^{1+2{\sf f}} \cdot l$ and again, $g(\bar{v})\neq g(\bar{v}')$. This
proves the first part of the claim. For the second part, let $t=0$ and $t'=1$. Then, $k \equiv 1~(\mod~p)$. As before,
this can only happen when $k=1$ and hence the second part of the claim holds.
\end{proof}

Hence, it is enough to show that the probability that this register is bad is low enough. The probability of measuring
$\ket{r_1,\ldots,r_n}$ equals $\frac{1}{2M^n} \cdot |\{ (t,\bar{a}) ~|~ F(t,\bar{a})=(r_1,\ldots,r_n) \}|$. Notice that
this probability is the same as the probability that $F(t,\bar{a})=(r_1,\ldots,r_n)$ for randomly chosen $t$ and
$\bar{a}$. Hence, we consider a randomly chosen $t$ and $\bar{a}$. If $t=0$, let
$\bar{a}'=(a_1+\frac{u_1-m}{p},a_2+u_2,\dots,a_n+u_n)$ and if $t=1$ let
$\bar{a}'=(a_1-\frac{u_1-m}{p},a_2-u_2,\dots,a_n-u_n)$.

\begin{claim}
With probability at least $1-\frac{1}{(n\log (2M))^{\sf f}}$, for randomly chosen $t$ and $\bar{a}$, $\bar{a}'$ is in
${\cal A}$ and $F(1-t, \bar{a}') = F(t,\bar{a})$.
\end{claim}
\begin{proof}
We assume that $t=0$, the proof for $t=1$ is similar. According to Lemma~\ref{svp_coordinates}, $|u_i| < 2^{2n}$.
Hence, unless there exists an $i$ for which $a_i < 2^{2n}$ or $a_i > M-2^{2n}$, $\bar{a}'$ is guaranteed to be in
${\cal A}$. This happens with probability at most $n2^{2n+1}/M$ because $\bar{a}$ is a random element of ${\cal A}$.

Notice that $f(1,\bar{a}') - f(0,\bar{a}) = \bar{u}$. Since $w_1,\ldots,w_n$ are randomly chosen, the probability that
$F(1-t, \bar{a}')$ and $F(t,\bar{a})$ differ on the $i$'th coordinate is at most
$$ \frac{|\langle \bar{u}, \bar{e}_i \rangle|}{\C{cub}n^{\half+2{\sf f}}\cdot l} \le
   \frac{|\langle \bar{u}, \bar{e}_i \rangle|}{\C{cub}n^{\half+2{\sf f}}\cdot \|\bar{u}\|}.$$
By the union bound, the probability that $F(1-t, \bar{a}') \neq F(t,\bar{a})$ is at most
$$ \frac{\sum_i |\langle \bar{u}, \bar{e}_i \rangle|}{\C{cub}n^{\half+2{\sf f}}\cdot \|\bar{u}\|} \le
   \frac{1}{\C{cub}n^{2{\sf f}}}$$
where we used the fact that the $l_1$ norm of a vector is at most $\sqrt{n}$ times its $l_2$ norm.

The sum of the two error probabilities $n\frac{2^{2n+1}}{M} + \frac{1}{\C{cub} n^{2{\sf f}}}$ is at most
$\frac{1}{(n\log (2M))^{\sf f}}$ for $\C{cub}$ large enough.
\end{proof}

This concludes the proof of Lemma~\ref{main_lattice_lemma}.
\end{proof}

\subsection{An Improved Algorithm}\label{sec_improved_alg}

In this section we complete the proof of Theorem~\ref{theorem_svp}. The algorithm we describe has many similarities
with the one in the previous section. The main difference is that it is based on $n$-dimensional balls instead of
cubes. The idea is to construct a ball of the right radius around lattice points and to show that if two lattice points
are close then the two balls have a large intersection while for any two far lattice points the balls do not intersect.
For technical reasons, we will assume in this section that the lattice is a subset of $\Z^n$. Any lattice with rational
points can be scaled so that it is a subset of $\Z^n$. We begin with some technical claims:

\begin{claim}\label{ball_intersection}
For any $R>0$, let $B_n$ be the ball of radius $R$ centered around the origin in $\R^n$ and let $B'_n=B_n+\bar{d}$ for
some vector $\bar{d}$ be a shifted ball. Then, the relative $n$-dimensional volume of their intersection is at least
$1-O(\sqrt{n} \| \bar{d} \|/R)$, i.e.,
$$ \frac{\vol(B_n \cap B'_n)}{\vol(B_n)} \ge 1-O(\sqrt{n} \| \bar{d} \| / R).$$
\end{claim}
\begin{proof}
Consider a point $\bar{x} \in \R^n$ such that $\langle \bar{x},\bar{d} \rangle / \|\bar{d}\| \ge \|\bar{d}\|/2$, i.e.,
a point which is closer to the center of $B'_n$ than to the center of $B_n$. Notice that $\bar{x} \in B_n$ implies
$\bar{x} \in B'_n$. In other words, the cap $C_n$ of $B_n$ given by all such points $\bar{x}$ is contained in $B_n \cap
B'_n$. By using a symmetric argument for points $\bar{x} \in \R^n$ such that $\langle \bar{x},\bar{d} \rangle /
\|\bar{d}\| < \|\bar{d}\|/2$ we get,
$$ \vol(B_n \cap B'_n) = 2\cdot \vol(C_n).$$
We can lower bound the volume of $C_n$ by half the volume of $B_n$ minus the volume of an $n$-dimensional cylinder of
radius $R$ and height $\|\bar{d}\|/2$:
$$ \vol(C_n) \ge \frac{1}{2} \vol(B_n) - \frac{\|\bar{d}\|}{2} \vol(B_{n-1})$$
where $B_{n-1}$ is the $n-1$-ball of radius $R$. We complete the proof by using the estimate $\vol(B_{n-1})/\vol(B_n) =
O(\sqrt{n}/R)$,
$$ \vol(C_n)/\vol(B_n) \ge \frac{1}{2} - O(\sqrt{n}\|\bar{d}\|/R ).$$
\end{proof}

In the algorithm we will actually represent the balls using points of a fine grid. Therefore, we would like to say that
the above claim still holds if we consider the number of grid points inside $B_n$, $B'_n$ and $B_n \cap B'_n$ instead
of their volumes. The following claim is more than enough for our needs:

\begin{claim}[Special case of Proposition 8.7 in \cite{MicciancioBook}]\label{lattice_points_volume}
Let $L$ be an integer and consider the scaled integer grid $\frac{1}{L}\Z^n$. Then, for any convex body $Q$ that
contains a ball of radius $r \ge \frac{1}{L}n^{1.5}$,
 $$ \left| \frac{|\frac{1}{L}\Z^n \cap Q|}{L^n \vol(Q)} - 1\right| < \frac{2n^{1.5}}{rL}.$$
\end{claim}

\begin{corollary}\label{ball_intersection_cor}
Let $L=2^n$ and consider the scaled integer grid $\frac{1}{L} \Z^n$. For any $R\ge 1$, let $B_n$ be the ball of radius
$R$ centered around the origin in $\R^n$ and let $B'_n=B_n+\bar{d}$ for some vector $\bar{d}$ such that $R/poly(n) \le
\|\bar{d}\| \le R$. Then, the relative number of grid points in their intersection is at least $1-O(\sqrt{n} \| \bar{d}
\|/R)$, i.e.,
$$ \frac{ | \frac{1}{L}\Z^n \cap B_n \cap B'_n | } { | \frac{1}{L}\Z^n \cap B_n | } \ge 1-O(\sqrt{n} \| \bar{d} \| / R).$$
\end{corollary}
\begin{proof}
We first note that $B_n$, $B'_n$ and $B_n \cap B'_n$ all contain the ball of radius $R/2 \ge 1/2$ centered around
$\bar{d}/2$. Using Claim \ref{lattice_points_volume} we obtain that the number of grid points in these bodies
approximates their volume up to a multiplicative error of $\frac{2n^{1.5}}{L/2}=2^{-\Omega(n)}$. We complete the proof
by using Claim \ref{ball_intersection}.
\end{proof}

Let $D(\cdot,\cdot)$ denote the trace distance between two quantum states \cite{NielsenChuang}. It is known that the
trace distance represents the maximum probability of distinguishing between the two states using quantum measurements.
We need the following simple bound on the trace distance:
\begin{claim}\label{trace_distance_tensor}
For all $k>0$ and density matrices $\sigma_1,\ldots,\sigma_k,\sigma'_1,\ldots,\sigma'_k$,
$$ D(\sigma_1\otimes \ldots \otimes \sigma_k, \sigma'_1\otimes \ldots \otimes \sigma'_k)
   \le \sum_{i=1}^k D(\sigma_i, \sigma'_i)$$
\end{claim}
\begin{proof}
Using the triangle inequality,
 \begin{eqnarray*}
  D(\sigma_1\otimes \ldots \otimes \sigma_k ,~ \sigma'_1\otimes \ldots \otimes \sigma'_k)
 &\le&D(\sigma_1\otimes \ldots \otimes \sigma_k,~
       \sigma'_1\otimes \sigma_2 \otimes \ldots \otimes \sigma_k) + \\
 &&D(\sigma'_1\otimes \sigma_2 \otimes \ldots \otimes \sigma_k,~
       \sigma'_1\otimes \sigma'_2 \otimes \sigma_3 \otimes \ldots \otimes \sigma_k) + \ldots \\
 &&D(\sigma'_1\otimes \ldots \otimes \sigma'_{k-1} \otimes \sigma_k,~
       \sigma'_1\otimes \ldots \otimes \sigma'_k)  \\
 &=& D(\sigma_1,~ \sigma'_1)+D(\sigma_2,~ \sigma'_2)+\ldots+D(\sigma_k,~ \sigma'_k).
 \end{eqnarray*}
\end{proof}

In addition, we will need the following lemma:
\begin{lemma}\label{quantum_superposition}
For any $1\le R \le 2^{poly(n)}$, let
$$ \ket{\eta} = \frac{1}{\sqrt{|\frac{1}{L}\Z^n \cap B_n|}}\sum_{\bar{x} \in \frac{1}{L}\Z^n \cap B_n} \ket{\bar{x}} $$
be the uniform superposition on grid points inside a ball of radius $R$ around the origin where $L=2^n$. Then, for any
$c>0$, a state $\ket{\tilde{\eta}}$ whose trace distance from $\ket{\eta}$ is at most $1/n^c$ can be efficiently
computed.
\end{lemma}
\begin{proof}
In order to bound the trace distance, we will use the fact that for any two pure states $\ket{\psi_1},\ket{\psi_2}$,
 \begin{equation}\label{equ:trace_distance}
 D(\ket{\psi_1},\ket{\psi_2}) = \sqrt{1-|\langle \psi_1 | \psi_2 \rangle|^2} \le \| \ket{\psi_1} - \ket{\psi_2} \|_2.
 \end{equation}
The first equality appears in \cite{NielsenChuang} and the inequality follows by a simple calculation.

Consider the (continuous) uniform probability distribution $q$ over $B_n$. Then one can define its discretization $q'$
to the grid $\frac{1}{L} \Z^n$ as
$$q'(\bar{x}) = \int_{\bar{x}+[0, 1/L]^n} q(\bar{y}) d\bar{y}$$
for $\bar{x} \in \frac{1}{L} \Z^n$. In other words, $q'(\bar{x})$ is proportional to the volume of the intersection of
$B_n$ with the cube $\bar{x}+[0,1/L]^n$. Notice that for points $\bar{x}$ such that $\bar{x}+[0,1/L]^n$ is completely
contained in $B_n$, $q'(\bar{x})=1/(L^n \vol(B_n))$. We claim that the state
$$ \ket{\eta'} = \sum_{\bar{x}\in \frac{1}{L} \Z^n}  \sqrt{q'(\bar{x})} \ket{\bar{x}} $$
is exponentially close to $\ket{\eta}$. Intuitively, this holds since the two differs only on points which are very
close to the boundary of the ball, namely, of distance $\sqrt{n}/L$ from the boundary. The number of such points is
negligible compared to the number of points in the interior of the ball. More formally, define
$$ \ket{\eta''} = \sqrt{\frac{L^n\vol(B_n)}{|\frac{1}{L}\Z^n \cap B_n|}} \ket{\eta'}.$$
Using Equation \ref{equ:trace_distance},
$$ D(\ket{\eta}, \ket{\eta'}) \le \| \ket{\eta'} - \ket{\eta} \|_2 \le \| \ket{\eta'} - \ket{\eta''} \|_2 + \| \ket{\eta''} - \ket{\eta} \|_2.$$
The first term is at most $2^{-\Omega(n)}$ according to Claim~\ref{lattice_points_volume}. For the second term, notice
that the amplitudes of $\ket{\eta''}$ and $\ket{\eta}$ are the same except possibly on points $\bar{x}$ of distance
$\sqrt{n}/L$ from the boundary. Using Claim~\ref{lattice_points_volume} again we get that the fraction of such points
is closely approximated by one minus the ratio of volumes of the ball of radius $R-\sqrt{n}/L$ and the ball of radius
$R$. This ratio of volumes is $(1- \sqrt{n} /(RL))^n \ge (1-\sqrt{n}/L)^n \ge 1-n^{1.5}/L = 1-2^{-\Omega(n)}$.

In the following we show how to approximate the state $\ket{\eta'}$. This idea is essentially due to Grover and Rudolph
\cite{GroverCreatingSuper}. Let $m\in \Z$ be large enough so that $B_n$ is contained in the cube $[-2^m, 2^m]^n$. Using
our assumption on $R$, $m<n^{c_1}$ for some $c_1 \ge 1$. We represent $\bar{x}$ using $K=n(m+1+\log L) < 2n^{1+c_1}$
qubits, i.e., a block of $m+1+\log L$ qubits for each dimension. Hence, we can write $\ket{\eta'}$ as
$$ \ket{\eta'} = \sum_{x_1,\ldots,x_K\in \{0,1\}}  \sqrt{q'(x_1,\ldots,x_K)} \ket{x_1,\ldots,x_K}.$$

We now show an equivalent way of writing $\ket{\eta'}$. Let us extend the definition of $q'$ in the following way: for
any $k\le K$ and any $x_1,\ldots, x_k \in \{0,1\}$ define $q'(x_1,\ldots,x_k)$ as the sum of
$q'(x_1,\ldots,x_k,x_{k+1},\ldots,x_K)$ over all sequences $x_{k+1},\ldots,x_K\in \{0,1\}$. Notice that
$q'(x_1,\ldots,x_k)$ corresponds to the volume of the intersection of $B_n$ with a certain cuboid (also known as a
rectangular parallelepiped). For example, $q'(0)=q'(1)=\frac{1}{2}$ since they represent the intersection of $B_n$ with
two halves of the cube $[-2^m,2^m]^n$. Using the definition $s(x_1)=q'(x_1)$ and for $k>1$,
$s(x_1,\ldots,x_k)=q'(x_1,\ldots,x_k)/q'(x_1,\ldots,x_{k-1})$ we see that
$$ \ket{\eta'} = \sum_{x_1 \in \{0,1\}} \sqrt{s(x_1)}
                 \sum_{x_2 \in \{0,1\}} \sqrt{s(x_1,x_2)} \ldots
                 \sum_{x_K \in \{0,1\}} \sqrt{s(x_1,\ldots,x_K)}
                 \ket{x_1,\ldots,x_K}.$$

The algorithm starts with all $K$ qubits in the state $\ket{0}$ and sets one qubit at a time. The first qubit is
rotated to the state $\frac{1}{\sqrt{2}} (\ket{0}+\ket{1})$. Assume we are now in the $k$'th step after setting the
state of qubits $1,\ldots,k-1$. We use the fact that there exists a classical algorithm for approximating the volume of
a convex body up to any $1/poly(n)$ error (see \cite{KannanLovaszVolume} and references therein). The body should be
provided by a ``well-guaranteed weak membership oracle", i.e., a sphere containing the body, a sphere contained in the
body, both of non-zero radius and an oracle that given a point decides if it is inside the body or not. It is easy to
construct such two spheres and an oracle for a body given by the intersection of a ball with a cuboid. Hence, we can
compute two values $\tilde{s}(x_1,\ldots,x_{k-1},0)$ and $\tilde{s}(x_1,\ldots,x_{k-1},1)$ such that
$$ \tilde{s}(x_1,\ldots,x_{k-1},0) + \tilde{s}(x_1,\ldots,x_{k-1},1) = 1$$
and
 $$ \left| \frac{\tilde{s}(x_1,\ldots,x_{k-1},i) }{ s(x_1,\ldots,x_{k-1},i)}  -  1  \right| < n^{-c_2}$$
for $i=0,1$ and some constant $c_2$ which will be chosen later. Then, we rotate the $i$'th qubit to the state
$\sqrt{\tilde{s}(x_1,\ldots,x_{k-1},0)} \ket{0} + \sqrt{\tilde{s}(x_1,\ldots,x_{k-1},1)} \ket{1}$. This completes the
description of the procedure.

Notice that the amplitude of each basis state $\ket{x_1,\ldots,x_K}$ in the resulting state $\ket{\tilde{\eta}}$ is
given by
$$ \prod_{k=1}^K \sqrt{\tilde{s}(x_1,\ldots,x_k)} \ge (1-n^{-c_2})^K \prod_{k=1}^K \sqrt{s(x_1,\ldots,x_k)}.$$
Hence the inner product $\langle \tilde{\eta} | \eta' \rangle$ is at least
 \begin{eqnarray*}
   (1-n^{-c_2})^K \sum_{x_1,\ldots,x_K \in \{0,1\}} \prod_{k=1}^K s(x_1,\ldots,x_k) &=&
   (1-n^{-c_2})^K \sum_{x_1,\ldots,x_K \in \{0,1\}} q'(x_1,\ldots,x_K) \\
   &=&  (1-n^{-c_2})^K \ge 1-K\cdot n^{-c_2} \ge 1-2 n^{1+c_1-c_2}.
 \end{eqnarray*}

Using Equation \ref{equ:trace_distance},
$$D(\ket{\eta'},\ket{\tilde{\eta}}) = \sqrt{1-|\langle \tilde{\eta} | \eta' \rangle|^2} < n^{-c}$$
for a large enough $c_2$.
\end{proof}

Let $p > n^{2+2{\sf f}}$ be any fixed prime. The following is the main lemma of this section. It essentially replaces
Lemma~\ref{main_lattice_lemma} and hence implies Theorem~\ref{theorem_svp}.

\begin{lemma}\label{improved_main_lattice_lemma}
For any ${\sf f}>0$ let $\bar{u}=\sum_{i=1}^{n} u_i \bar{b}_i$ be the shortest lattice vector in a $(\C{unq}
n^{\frac{1}{2}+2{\sf f}})$-unique lattice where $\C{unq}>0$ is a constant. If there exists a solution to the two point
problem with failure parameter ${\sf f}$ then there exists a quantum algorithm that given this lattice and three
integers $l,m,i_0$ returns $(u_1,\ldots,u_{i_0-1},\frac{u_{i_0}-m}{p},u_{i_0+1},\ldots,u_n)$ with probability
$1/poly(n)$ if the following conditions hold: $\|\bar{u}\| \le l \le 2\|\bar{u}\|$, $u_{i_0}\equiv m~(\mod~p)$ and
$1\le m\le p-1$.
\end{lemma}
\begin{proof}
As before, let $\langle \bar{b}_1,\ldots,\bar{b}_n \rangle$ be an LLL reduced basis, let $M=2^{4n}$ and assume that
$i_0=1$. We also define $f(t,\bar{a})$ as before. Assume that the number of registers needed by the two point algorithm
is at most $n^{c_1}$ for some constant $c_1>0$.

The algorithm starts by calling the routine of Claim \ref{quantum_superposition} $n^{c_1}$ times with accuracy
parameter $n^{-c_2}$ and $R=\C{bal}n^{\frac{1}{2}+2{\sf f}}\cdot l$ for some constants $c_2,\C{bal}>0$. The state we
obtain is
 \begin{equation}\label{equ_approx_state}
 \ket{\tilde{\eta}_1} \otimes \ldots \otimes \ket{\tilde{\eta}_{n^{c_1}}}
 \end{equation}
where each $\ket{\tilde{\eta}_i}$ has a trace distance of at most $n^{-c_2}$ from $\ket{\eta}$. According to
Claim~\ref{trace_distance_tensor}, the above tensor product has a trace distance of at most $n^{c_1-c_2}$ from
$\ket{\eta}^{\otimes n^{c_1}}$. In the following we show that the algorithm succeeds with probability at least
$n^{-c_3}$ for some $c_3>0$ given the state $\ket{\eta}^{\otimes n^{c_1}}$. This would complete the proof since given
the state in Equation \ref{equ_approx_state}, the algorithm succeeds with probability at least
$n^{-c_3}-n^{c_1-c_2}>\frac{1}{2}n^{-c_3}$ for large enough $c_2$.

We describe a routine that given the state $\ket{\eta}$ creates one register in the input to the two point problem. In
order to produce a complete input to the two point problem, the algorithm calls this routine $n^{c_1}$ times, each time
with a new $\ket{\eta}$ register. It then calls the two point algorithm and outputs the result. As required, the
success probability is $1/poly(n\log M)=n^{-c_3}$ for some $c_3>0$.

Given $\ket{\eta}$, the routine creates the state
 $$ \frac{1}{\sqrt{2M^n}}\sum_{t\in \{0,1\}, \bar{a} \in {\cal A}} \ket{t,\bar{a}} \otimes \ket{\eta},$$
or equivalently,
 $$ \sum_{t\in \{0,1\}, \bar{a} \in {\cal A}, \bar{x} \in \frac{1}{L}\Z^n \cap B_n} \ket{t,\bar{a},\bar{x}} $$
where $B_n$ is the ball of radius $R$ around the origin and $L=2^n$. We add the value $f(t,\bar{a})$ to the last
register,
 $$ \sum_{t\in \{0,1\}, \bar{a} \in {\cal A}, \bar{x} \in \frac{1}{L}\Z^n \cap B_n} \ket{t,\bar{a},f(t,\bar{a})+\bar{x}}. $$
Finally, we measure the last register and if $\bar{x}'$ denotes the result, the state collapses to
 $$ \sum_{t\in \{0,1\}, \bar{a} \in {\cal A} | \bar{x}' \in f(t,\bar{a})+\frac{1}{L}\Z^n \cap B_n}
     \ket{t,\bar{a},\bar{x}'}. $$

\begin{claim}
For every $\bar{x}'$, there is at most one element of the form $(0,\bar{a})$ and at most one element of the form
$(1,\bar{a}')$ such that $\bar{x}' \in f(t,\bar{a})+\frac{1}{L}\Z^n \cap B_n$. Moreover, if there are two such elements
$(0,\bar{a})$ and $(1,\bar{a}')$ then $\bar{a}'-\bar{a}$ is the vector $(\frac{u_1-m}{p},u_2,\ldots,u_m)$.
\end{claim}
\begin{proof}
Consider two different lattice points in the image of $f$, $\bar{v}=f(t,\bar{a})$ and $\bar{v}'=f(t',\bar{a}')$, such
that $\bar{x}'$ is both in $\bar{v}+\frac{1}{L}\Z^n \cap B_n$ and $\bar{v}'+\frac{1}{L}\Z^n \cap B_n$. This implies
that $\|\bar{v}-\bar{v}'\| \le \C{bal}n^{\frac{1}{2}+2{\sf f}}\cdot l \le 2\C{bal}n^{\frac{1}{2}+2{\sf f}}\cdot
\|\bar{u}\|$. For $\C{unq} > 2\C{bal}$ this means that $\bar{v}'-\bar{v}=k \cdot \bar{u}$ for some integer $k\neq 0$.
As before, by considering the first coordinate of $\bar{v}'-\bar{v}$ in the lattice basis we get that $(a'_1 p +
t'm)-(a_1 p + tm) \equiv k\cdot m~(\mod~p)$. Hence, $k \equiv t'-t~(\mod~p)$. If $t=t'$ then $k\equiv 0~(\mod~p)$ and
therefore $|k| \ge p$ which contradicts the above upper bound on the distance between $\bar{v}$ and $\bar{v}'$. This
proves the first part of the claim. For the second part, let $t=0$ and $t'=1$. Then, $k \equiv 1~(\mod~p)$. As before,
this can only happen when $k=1$ and hence the second part of the claim holds.
\end{proof}

Notice that the probability of measuring $\bar{x}'$ is the same as that obtained by first choosing random $t$ and
$\bar{a}$ and then choosing a random point in $f(t,\bar{a})+\frac{1}{L}\Z^n \cap B_n$. Let us define for any $t$ and
$\bar{a}$ the vector $\bar{a}'$ as before.

\begin{claim}
With probability at least $1-\frac{1}{(n\log (2M))^{\sf f}}$, for randomly chosen $t$ and $\bar{a}$ and a random point
$\bar{x}'$ in $f(t,\bar{a})+\frac{1}{L}\Z^n \cap B_n$, $\bar{a}'$ is in ${\cal A}$ and $\bar{x}'$ is also in
$f(1-t,\bar{a}')+\frac{1}{L}\Z^n \cap B_n$.
\end{claim}
\begin{proof}
According to Lemma~\ref{svp_coordinates}, $|u_i| < 2^{2n}$. Hence, unless there exists an $i$ for which $a_i < 2^{2n}$
or $a_i > M-2^{2n}$, $\bar{a}'$ is guaranteed to be in ${\cal A}$. This happens with probability at most $n2^{2n+1}/M$
because $\bar{a}$ is a random element of ${\cal A}$.

Fix $\bar{a},\bar{a}' \in {\cal A}$. We would like to show that if $\bar{x}'$ is chosen uniformly from
$f(t,\bar{a})+\frac{1}{L}\Z^n \cap B_n$ then with high probability it is also in $f(1-t,\bar{a}')+\frac{1}{L}\Z^n \cap
B_n$. By translating both sets by $-f(t,\bar{a})$ we get the equivalent statement that if $\bar{x}'$ is chosen
uniformly from $\frac{1}{L}\Z^n \cap B_n$ then with high probability it is also in
$(f(1-t,\bar{a}')-f(t,\bar{a}))+\frac{1}{L}\Z^n \cap B_n$. Since we assumed that our lattice is a subset of $\Z^n$,
$f(1-t,\bar{a}')-f(t,\bar{a}) \in \Z^n$ and the latter set equals $\frac{1}{L}\Z^n \cap
(f(1-t,\bar{a}')-f(t,\bar{a})+B_n)$. Using Corollary \ref{ball_intersection_cor} and the fact that
$\|f(1-t,\bar{a}')-f(t,\bar{a})\| = \|\bar{u}\| \le l$, we get that the required probability is at least
$$1-O(\sqrt{n} l / R) = 1-O(\sqrt{n} l / (\C{bal}n^{\frac{1}{2}+2{\sf f}} \cdot l))
   = 1-O(1 / (\C{bal}n^{2{\sf f}})).$$

The sum of the two error probabilities $n\frac{2^{2n+1}}{M} + O(1 / (\C{bal}n^{2{\sf f}}))$ is at most $\frac{1}{(n\log
(2M))^{\sf f}}$ for $\C{bal}$ large enough.
\end{proof}

This concludes the proof of Lemma~\ref{improved_main_lattice_lemma}.
\end{proof}

\section{The Dihedral Coset Problem}
\label{section_two_point}

\alternate{We begin this section with a description of the average case subset sum problem. We describe our assumptions on the
subroutine that solves it and prove some properties of such a subroutine. In the second subsection we present an
algorithm that solves the DCP with calls to an average case subset sum subroutine.}{}

\subsection{Subset Sum}

The subset sum problem is defined as follows. An input is a sequence of numbers $A=(a_1,\ldots,a_r)$ and two numbers
$t,N$. The output is a subset $B\subseteq [r]$ such that $\sum_{i \in B} a_i \equiv t~(\mod~N)$. Let a legal input be
an input for which there exists a subset $B$ with $\sum_{i \in B} a_i \equiv t~(\mod~N)$. For a constant $\C{r}>0$, we
fix $r$ to be $\log N + \C{r}$ since we will only be interested in such instances. \alternate{First we show that there
are many legal inputs:

\begin{lemma}\label{many_legal}
For randomly chosen $a_1,\ldots,a_r,t$ in $\{0,\ldots,N-1\}$, the probability that there is no $B\subseteq [r]$ such
that $\sum_{i \in B} a_i \equiv t~(\mod~N)$ is at most $\frac{1}{2}$.
\end{lemma}
\begin{proof}
Fix a value of $t$. Define a random variable $X_{\bar{b}}$ for every $\bar{b} \in \{0,1\}^r, \bar{b}\neq 0^r$ as $1$ if
$\sum_i b_i a_i \equiv t~(\mod~N)$ and $0$ otherwise. Since for every $\bar{b}$ the sum $\sum_{i} b_i a_i$ has any
value modulo $N$ with the same probability, the expectation of $X_{\bar{b}}$ is $\frac{1}{N}$ and its variance is
$\frac{1}{N} - \frac{1}{N^2}<\frac{1}{N}$. Hence,
$$E[\sum_{\bar{b}} X_{\bar{b}} ] = \sum_{\bar{b}} E[X_{\bar{b}}] = \frac{2^r-1}{N}$$
Given two different sequences $\bar{b},\bar{b}' \in \{0,1\}^r$ we show that $X_{\bar{b}}$ and $X_{\bar{b}'}$ are
independent. Let $i$ be such that $b_i \neq b'_i$ and assume without loss of generality that $b_i = 1, b'_i = 0$ and
$i=1$. Then,
\begin{eqnarray*}
Pr_{a_1,\ldots,a_r}[X_{\bar{b}} = 1~\wedge~X_{\bar{b}'} = 1] &=&
E_{a_2,\ldots,a_r}[Pr_{a_1}[X_{\bar{b}} = 1~\wedge~X_{\bar{b}'} = 1]]  \\
&=&E_{a_2,\ldots,a_r}[1/N\cdot \delta_{X_{\bar{b}'},1}] \\
&=&Pr_{a_1,\ldots,a_r}[X_{\bar{b}} = 1]Pr_{a_1,\ldots,a_r}[X_{\bar{b}'} = 1]
\end{eqnarray*}
where the second equality holds because $X_{\bar{b}'}$ does not depend on $a_1$ and $X_{\bar{b}}$ is 1 with probability
$1/N$ for any $a_2,\ldots,a_r$. A similar argument holds for other values of $X_{\bar{b}}$ and $X_{\bar{b}'}$.
Therefore, the random variables are pairwise independent and by the Chebyshev bound,
$$Pr[\sum_{\bar{b}} X_{\bar{b}} < \frac{1}{2} \cdot \frac{2^r-1}{N}] \le 4\cdot\frac{N}{2^r-1} \le \frac{8}{2^{\C{r}}}.$$
In particular, the probability of $\sum_{\bar{b}} X_{\bar{b}} = 0$, that is, the probability that there is no $B$ such
that $\sum_{i \in B} a_i \equiv t~(\mod~N)$ is at most $\frac{8}{2^{\C{r}}}=\frac{1}{2}$ for $\C{r}=4$.
\end{proof}

We assume that we are given a subroutine that answers a $\frac{1}{\log ^ {\C{s}} N}$ fraction of the legal subset sum
inputs with parameter $N$ where $\C{s}>0$ is any constant. As can be seen from the previous lemma, this implies that
the subroutine answers a non-negligible fraction of all inputs (and not just the legal inputs). In addition, we
assume that the subroutine is deterministic. 
}{We assume that we are given a deterministic subroutine that answers a $\frac{1}{\log ^ {\C{s}} N}$ fraction of the
legal subset sum inputs with parameter $N$ where $\C{s}>0$ is any constant.}We denote by $S(A,t)$ the result of the
subroutine $S$ on the input $A=(a_1,\ldots,a_r),t$ and we omit $N$. This result can either be a set or an error. Let
$S(A)$ denote the set of $t$'s for which the subroutine returns a set and not an error, i.e., $S(A)=\{t~|~S(A,t) \neq
error\}$.

\alternate{\begin{corollary}\label{big_sa}
For randomly chosen $a_1,\ldots,a_r$ in $\{0,\ldots,N-1\}$, $Pr_A[|S(A)| \ge \frac{N}{4\log^{\C{s}} N}] =
\Omega(\frac{1}{\log^{\C{s}} N})$ where $A=(a_1,\ldots,a_r)$.
\end{corollary}
\begin{proof}
Since $S(A,t)\neq error$ only when $(A,t)$ is a legal input,
\begin{eqnarray*}
Pr_{A,t}[S(A,t)\neq error] &=& Pr_{A,t}[~S(A,t)\neq error ~\wedge~ (A,t)~\mbox{is legal}~] \\&=& Pr_{A,t}[~S(A,t)\neq
error ~|~ (A,t)~\mbox{is legal}~] \cdot Pr_{A,t}[~(A,t)~\mbox{is legal}~] \ge \frac{1}{2\log^{\C{s}}N}.
\end{eqnarray*}
In addition,
\begin{eqnarray*}
Pr_{A,t}[S(A,t)\neq error] &=& E_A[~\frac{|S(A)|}{N}~] \\&\le&
   Pr_A[~|S(A)|\ge \frac{N}{4\log^{\C{s}}N}~] + Pr_A[~|S(A)| < \frac{N}{4\log^{\C{s}}N}~] \cdot \frac{1}{4\log^{\C{s}}N}
   \\&\le&
   Pr_A[~|S(A)|\ge \frac{N}{4\log^{\C{s}}N}~] + \frac{1}{4\log^{\C{s}}N}.
\end{eqnarray*}
By combining the two inequalities we obtain the corollary.
\end{proof}

\begin{lemma}\label{finding_q_prime}
Let $T\subseteq \{0,\ldots,N-1\}$ be a set such that $|T| > \frac{N}{s}$ for a certain $s$. Then, for any
$q<\frac{N}{8s}$ there exists $q'\in\{q,2q,\ldots,sq\}$ such that the number of pairs $t,t+q'$ that are both in $T$ is
$\Omega(\frac{N}{s^3})$.
\end{lemma}
\begin{proof}
Define the partition of $T$ into sets $T_0,\ldots,T_{q-1}$ as
$$T_k=\{i~|~i\in T,i \equiv k~(\mod~q)\}.$$
At least $\frac{q}{2s}$ of the sets are of size at least $\frac{N}{2sq}$ since their union is $T$ and
$\frac{q}{2s}\frac{N}{q} + \frac{N}{2s} < |T|$. Let $T_i$ be such a set and for $t\in T_i$ consider the values
$t+q,t+2q,\ldots,t+4sq$. Therefore, the number of $t \in T_i$ such that none of these values is in $T_i$ is less than
$\frac{N}{4sq}$ because $|\{i ~|~ 0\le i< N, i \equiv k~(\mod~q)\}| = \frac{N}{q}$. Therefore, more than
$|T_i|-\frac{N}{4sq}\ge \frac{N}{4sq}$ of the elements $t\in T_i$ are such that one of $t+q,t+2q,\ldots,t+4sq$ is also
in $T_i$. Summing over all sets $T_i$ such that $|T_i|\ge \frac{N}{2sq}$, there are at least $\frac{N}{4sq} \cdot
\frac{q}{2s} = \frac{N}{8s^2}$ elements $t\in T$ for which one of $t+q,t+2q,\ldots,t+4sq$ is also in $T$. Thus, there
exists a $q' \in \{ q, 2q, \ldots, 4sq \}$ such that the number of $t\in T$ for which $t+q'\in T$ is at least
$\frac{N}{32s^3}$.
\end{proof}

}{}\begin{definition}
A partial function $f:\{0,\ldots,N-1\} \rightarrow \{0,\ldots,N-1\}$ is called a matching if for all $i$ such that
$f(i)$ is defined, $f(i) \neq i$ and $f(f(i))=i$. A matching is a $q$-matching if for all $i$ such that $f(i)$ is
defined, $|f(i)-i|=q$. We define an equal partition of the domain of a matching $f$ by $A_1(f)=\{i ~|~ f(i)~defined
~\wedge~ f(i)>i\}$ and $A_2(f)=\{i ~|~ f(i)~defined ~\wedge~ f(i)<i\}$. The intersection of a matching $f$ and a set $T
\subseteq \{0,\ldots,N-1\}$ is the set $\{i ~|~ i\in T ~\wedge~ f(i) \in T\}$.
\end{definition}

For any $q$ we define the following $q$-matchings:

$$
f^1_{q}(t)= \left\{
\begin{array}{ll}
    t+q & t~\mod~ 2q < q,~ t+q<N, \\
    t-q & t~\mod~ 2q \ge q,~ t-q\ge 0, \\
    undefined & \mbox{otherwise}.
\end{array} \right.
~ f^2_{q}(t)= \left\{
  \begin{array}{ll}
    t-q & t~\mod~ 2q < q,~ t-q\ge 0,\\
    t+q & t~\mod~ 2q \ge q,~ t+q < N, \\
    undefined & \mbox{otherwise}.
  \end{array} \right.
$$

\begin{lemma}\label{lemma_matching}
There exists a constant $\C{m}$ such that for any integer $q < \frac{N}{\log^{\C{m}} N}$ there exists a matching $f$
among the $2\log^{\C{m}} N$ matchings $f^1_q, f^1_{2q}, \ldots, f^1_{\log^{\C{m}} N q},f^2_q, f^2_{2q}, \ldots,
f^2_{\log^{\C{m}} N q}$ such that with probability at least $\frac{1}{\log^{\C{m}} N}$ on the choice of $A$, the
intersection of $f$ and $S(A)$ is $\frac{N}{\log^{\C{m}} N}$. We call such an $f$ a {\em good} matching.
\end{lemma}
\begin{proof}\omittedproof{
According to Corollary~\ref{big_sa}, $\frac{1}{4\log^{\C{s}} N}$ of the possible values of $A$ satisfy $|S(A)| >
\frac{N}{4\log^{\C{s}} N}$. For such $A$, Lemma~\ref{finding_q_prime} with $s=4\log^{\C{s}} N$ implies that there
exists a value $q'\in\{q,2q,\ldots,4\log^{\C{s}} N \cdot q \}$ such that the number of pairs $t,t+q'$ that are both in
$S(A)$ is $\Omega(\frac{N}{\log^{3\C{s}} N})$. Therefore, for such $A$ and $q'$, the size of the intersection of one of
the matchings $f^1_{q'}, f^2_{q'}$ and $S(A)$ is $\Omega(\frac{N}{\log^{3\C{s}} N})$. This implies that one of the
$8\log^{\C{s}} N$ matchings considered must have an intersection of size $\Omega(\frac{N}{\log^{3\C{s}} N})$ with at
least $\frac{1}{32\log^{2\C{s}} N}$ of the possible values of $A$. We conclude the proof by choosing $\C{m}>3\C{s}$.
}\end{proof}

\subsection{The Quantum Algorithm}

\alternate{We begin with the following simple claim:}{}

\begin{claim}\label{to_one_qubit}
For any two basis states $\ket{a}$ and $\ket{b}$, $a\neq b$, there exists a routine such that
given the state $\ket{a}+e(\phi)\ket{b}$ outputs the state $\ket{0}+e(\phi)\ket{1}$.
\end{claim}
\begin{proof}
Consider the function $f$ defined as $f(a)=0, f(0)=a, f(b)=1, f(1)=b$ and $f(i)=i$ otherwise. It is reversible and can
therefore be implemented as a quantum routine.
\end{proof}

We now describe the main routine in the DCP algorithm.

\begin{lemma}\label{routine_a}
There exist routines $R_1,R_2$ such that given a $q$-matching $f$ and an input for the DCP with failure parameter $1$,
they either output a bit or they fail. Conditioned on non-failure, the probability of the bit being 1 is
$\frac{1}{2}-\frac{1}{2}\cos(2\pi q \frac{d}{N})$ for $R_1$ and $\frac{1}{2}+\frac{1}{2}\sin(2\pi q \frac{d}{N})$ for
$R_2$. Moreover, if $f$ is a good matching, the success probability is $\Omega(\frac{1}{\log^{\C{m}}N})$.
\end{lemma}
\begin{proof}
The routines begin by performing a Fourier transform on the last $\log N$ qubits of each input register. Consider one
register. Assuming it is a good register, the resulting state is \alternate{
\begin{eqnarray*}
&&\frac{1}{\sqrt{2N}}\sum_{i=0}^{N-1}e(ix/N)\ket{0,i} + \frac{1}{\sqrt{2N}}\sum_{i=0}^{N-1}e(i(x+d)/N)\ket{1,i} =\\
&&\frac{1}{\sqrt{2N}}\sum_{i=0}^{N-1}e(ix/N)(\ket{0} + e(id/N)\ket{1})\ket{i} .
\end{eqnarray*}}{
$$\frac{1}{\sqrt{2N}}\sum_{i=0}^{N-1}e(ix/N)(\ket{0} + e(id/N)\ket{1})\ket{i} .$$
}
We measure the last $\log N$ qubits and let $a \in \{0,\ldots,N-1\}$ be the result. The state collapses to
$$ \frac{1}{\sqrt{2}}e(ax/N)(\ket{0} + e(ad/N)\ket{1})\ket{a} .$$
If it is a bad register, it is in the state $\ket{b,x}$ where both $b$ and $x$ are arbitrary. After the Fourier
transform the state is $\frac{1}{\sqrt{N}}\sum_{i=0}^{N-1}e(ix/N)\ket{b,i}$ and after measuring $a$ in the last $\log
N$ qubits, the state is $e(ax/N)\ket{b,a}$. Notice that in both cases any value $a$ in $\{0,\ldots,N-1\}$ has an equal
probability of being measured.

We choose the number of input registers to be $r$. Let $A=(a_1,\ldots,a_r)$ be the sequence of values measured in the
above process. Notice that this sequence is uniform and hence can be used as an input to the average case subset sum
algorithm. In the following, we assume that $s$ of the $r$ registers are bad. Later we will claim that with good
probability, none of the registers is bad. Yet, we have to show that even if one of the registers is bad, the routine
does not return erroneous results. Without loss of generality, assume that the first $s$ registers are bad. The
resulting state is\alternate{:
$$\bigotimes_{i=1}^{s}[e(a_i x_i/N)\ket{b_i,a_i}]\bigotimes_{i=s+1}^{r}[\frac{1}{\sqrt{2}}e(a_i x_i/N)(\ket{0} +
e(a_id/N)\ket{1})\ket{a_i}].$$ Or, by omitting the multiplication by the fixed phase and the $r\cdot \lceil\log
N\rceil$ fixed qubits,}{ (omitting a fixed phase and the fixed qubits): }
$$\bigotimes_{i=1}^{s}[\ket{b_i}]\bigotimes_{i=s+1}^{r}[\frac{1}{\sqrt{2}}(\ket{0} + e(a_id/N)\ket{1})].$$ Denote
these $r$ qubits by $\balpha = (\alpha_1,\ldots,\alpha_r)$.

We add $r+1$ new qubits, $\bbeta = (\beta_1,\ldots,\beta_r)$ and $\gamma$. Let $t_{\balpha}$ denote the sum $\sum_{i=1}^r
\alpha_i a_i$. Next, we perform the following operations:

\begin{center}
\alternate{
\fbox{
 \begin{minipage}{3in}
 \vspace{.1in}
 \begin{pseudocode}[display]{TwoPointRoutine}{f} \label{two_point_routine}
    \IF S(A,t_{\balpha})\neq \balpha ~\vee~ S(A,f(t_{\balpha}))=error \THEN exit \\
    \IF t_{\balpha}\in A_1(f) \THEN
    \BEGIN
        \bbeta \GETS \balpha \\
        \gamma \GETS 1
    \END
    \ELSEIF t_{\balpha}\in A_2(f) \THEN
    \BEGIN
        \bbeta \GETS S(A,f(t_{\balpha})) \\
        \gamma \GETS 1
    \END
    \ELSE
        exit
 \end{pseudocode}
 \vspace{.1in}
 \end{minipage}
}}{\fbox{
 \begin{minipage}{4in}
 \vspace{.1in}
 \begin{pseudocode}[display]{TwoPointRoutine}{f} \label{two_point_routine}
    \IF S(A,t_{\balpha})\neq \balpha ~\vee~ S(A,f(t_{\balpha}))=error~\mbox{\bf then exit} \\
    \IF t_{\balpha}\in A_1(f) \THEN
    \BEGIN
        \bbeta \GETS \balpha \\
        \gamma \GETS 1
    \END
    \ELSEIF t_{\balpha}\in A_2(f) \THEN
    \BEGIN
        \bbeta \GETS S(A,f(t_{\balpha})) \\
        \gamma \GETS 1
    \END
    \ELSE
        exit
 \end{pseudocode}
 \vspace{.1in}
 \end{minipage}
}}
\end{center}

In order to describe the state after the above procedure, we define the following subsets of $\{0,1\}^r$:
$$ M=\{ \balpha \in \{0,1\}^r ~|~ \alpha_1=b_1,\ldots,\alpha_s=b_s\} $$
$$ L=\{ \balpha \in M ~|~ t_{\balpha} \in A_1(f) ~\wedge~ S(A,t_{\balpha})=\balpha ~\wedge S(A,f(t_{\balpha}))\neq error \} $$
$$ R=\{ \balpha \in M ~|~ t_{\balpha} \in A_2(f) ~\wedge~ S(A,t_{\balpha})=\balpha ~\wedge S(A,f(t_{\balpha}))\neq error \} $$
Using the order $\ket{\balpha,\bbeta,\gamma}$, the resulting state is:
\alternate{\begin{eqnarray*}
&&\frac{1}{\sqrt{2^{r-s}}}(
  \sum_{\balpha \in M-L-R} e(\langle\balpha,\bar{a}\rangle \frac{d}{N} ) \ket{\balpha, \bar{0},0} +\\
&& \qquad  \sum_{\balpha \in L} e(\langle\balpha,\bar{a}\rangle \frac{d}{N} ) \ket{\balpha,\balpha,1} +
  \sum_{\balpha \in R} e(\langle\balpha,\bar{a}\rangle \frac{d}{N} )\ket{\balpha,S(A,f(t_{\balpha})),1})\\
&=& \frac{1}{\sqrt{2^{r-s}}}(
  \sum_{\balpha \in M-L-R} e(\langle\balpha,\bar{a}\rangle \frac{d}{N} ) \ket{\balpha, \bar{0},0} +\\
&&\qquad  \sum_{\balpha \in L} (e(\langle\balpha,\bar{a}\rangle \frac{d}{N} )\ket{\balpha,\balpha,1} +
                       e(\langle S(A,f(t_{\balpha})),\bar{a}\rangle \frac{d}{N} )\ket{S(A,f(t_{\balpha})),\balpha,1}))\\
&=& \frac{1}{\sqrt{2^{r-s}}}(
  \sum_{\balpha \in M-L-R} e(\langle\balpha,\bar{a}\rangle \frac{d}{N} ) \ket{\balpha, \bar{0},0} +\\
&&\qquad  \sum_{\balpha \in L} e(\langle\balpha,\bar{a}\rangle \frac{d}{N} ) ( \ket{\balpha}+ e(q\cdot \frac{d}{N})
\ket{S(A,f(t_{\balpha}))})\ket{\balpha, 1})
\end{eqnarray*}}{\begin{eqnarray*}
&&\frac{1}{\sqrt{2^{r-s}}}(
  \sum_{\balpha \in M-L-R} e(\langle\balpha,\bar{a}\rangle \frac{d}{N} ) \ket{\balpha, \bar{0},0} +
  \sum_{\balpha \in L} e(\langle\balpha,\bar{a}\rangle \frac{d}{N} ) \ket{\balpha,\balpha,1} +
  \sum_{\balpha \in R} e(\langle\balpha,\bar{a}\rangle \frac{d}{N} )\ket{\balpha,S(A,f(t_{\balpha})),1})\\
&=& \frac{1}{\sqrt{2^{r-s}}}(
  \sum_{\balpha \in M-L-R} e(\langle\balpha,\bar{a}\rangle \frac{d}{N} ) \ket{\balpha, \bar{0},0} +\\
&&\qquad  \sum_{\balpha \in L} (e(\langle\balpha,\bar{a}\rangle \frac{d}{N} )\ket{\balpha,\balpha,1} +
                       e(\langle S(A,f(t_{\balpha})),\bar{a}\rangle \frac{d}{N} )\ket{S(A,f(t_{\balpha})),\balpha,1}))\\
&=& \frac{1}{\sqrt{2^{r-s}}}(
  \sum_{\balpha \in M-L-R} e(\langle\balpha,\bar{a}\rangle \frac{d}{N} ) \ket{\balpha, \bar{0},0} +
  \sum_{\balpha \in L} e(\langle\balpha,\bar{a}\rangle \frac{d}{N} ) ( \ket{\balpha}+ e(q\cdot \frac{d}{N})
\ket{S(A,f(t_{\balpha}))})\ket{\balpha, 1})
\end{eqnarray*}}

Now we measure $\bbeta$ and $\gamma$. If $\gamma=0$, the routine failed. Otherwise, the state of $\balpha$ is (omitting
the fixed $\bbeta$ and $\gamma$):
$$\frac{1}{\sqrt{2}}(\ket{\bbeta}+e(q \cdot \frac{d}{N})\ket{S(A,f(t_{\bbeta}))}).$$
Notice that since $\bbeta$ is known and $S(A,f(t_{\bbeta}))$ can be easily found by calling $S$, we can transform this
state to the state
$$\frac{1}{\sqrt{2}}(\ket{0}+e(q \cdot \frac{d}{N})\ket{1})$$
by using Claim~\ref{to_one_qubit}. \alternate{By omitting some qubits, we can assume that this is a state on one qubit. By using
the Hadamard transform the state becomes
$$\frac{1}{2}((1+e(q \frac{d}{N}))\ket{0}+(1-e(q \frac{d}{N}))\ket{1}).$$
We measure the qubit and the probability of measuring $1$ is
$$\frac{1}{4}|1-e(q \frac{d}{N})|^2 = \frac{1}{4}(2-2\cos(2\pi q \frac{d}{N})) = \frac{1}{2} - \frac{1}{2} \cos(2\pi q\frac{d}{N}).$$
This completes the description of $R_1$. The routine $R_2$ applies the transform
$$ \left( \begin{array}{cc}
1 & 0 \\
0 & \imath \end{array} \right)
$$
before the Hadamard transform and thus the state becomes
$$\frac{1}{2}((1+e(1/4 + q \frac{d}{N}))\ket{0}+(1-e(1/4 + q \frac{d}{N}))\ket{1})$$
and the probability of measuring $1$ becomes $\frac{1}{2} - \frac{1}{2} \cos(\pi/2 + 2\pi q\frac{d}{N})=\frac{1}{2} +
\frac{1}{2} \sin(2\pi q\frac{d}{N})$.}{Routine $R_1$ applies the Hadamard transform and then measures the qubit. Notice
that the probability of measuring $1$ is $\frac{1}{2} - \frac{1}{2} \cos(2\pi q\frac{d}{N})$. Routine $R_2$ adds the
transform $ \left( \begin{array}{cc}
1 & 0 \\
0 & \imath \end{array} \right)
$ before the Hadamard transform.}

From the previous description, it is clear that the probability of measuring 1 conditioned on a non-failure is correct.
Thus, it remains to prove that when $f$ is a good matching the failure probability is low. The success probability
equals the probability of measuring $\gamma=1$ which is $|L\cup R|/2^{r-s}$. Assume that none of the $r$ registers is
bad. Then, $|L\cup R|/2^{r-s} = |L\cup R|/2^{r}$ and $L\cup R$ becomes $\{\balpha\in \{0,1\}^r ~|~ t_{\balpha} \in
A_1(f) \cup A_2(f) ~\wedge~ S(A,t_{\balpha})=\balpha ~\wedge~ S(A,f(t_{\balpha})) \neq error \}$. Notice that the size
of this set equals $|\{t ~|~ t\in S(A) ~\wedge~ f(t) \in S(A)\}|$ which, according to the definition of a good
matching, is at least $\frac{N}{\log ^{\C{m}}N}$. Therefore the probability of success conditioned on all of the
registers being good is $|L\cup R|/2^r = \frac{1}{2^{\C{r}}\log^{\C{m}} N} = \Omega(\frac{1}{\log^{\C{m}} N})$. This
concludes the proof since with probability at least $(1-\frac{1}{\log N})^r = (1-\frac{1}{\log N})^{\log N + \C{r}} =
\Omega(1)$ none of the registers is bad.
\end{proof}

\begin{claim}\label{sincos_estimate}
Given an approximation $x$ of $\sin\phi$ and an approximation $y$ of $\cos\phi$ with additive error $\epsilon$, we can
find $\phi~\mod~2\pi$ up to an additive error of $O(\epsilon)$.
\end{claim}
\begin{proof}\omittedproof{
Assume $y \ge 0$ and let $z=\frac{x}{1+y}$. A simple calculation shows that $z$ is an estimate of
$\frac{\sin\phi}{1+\cos\phi}$ up to an additive error of at most $4\epsilon$. The estimate on $\phi$ is $2\arctan z$.
Since the absolute value of the differential of $\arctan$ is at most $1$, this is an estimate of
$2\arctan(\frac{\sin\phi}{1+\cos\phi})=\phi$ with an additive error of at most $8\epsilon$. When $y < 0$ we compute an
estimate of  $2\mbox{arccot}(\frac{\sin\phi}{1-\cos\phi})=\phi$. }
\end{proof}

\begin{lemma}
There exists a routine $R_3$ such that with probability exponentially close to $1$, given any $q < \frac{N}{\log^{\C{m}}
N}$ finds a value $q' \in \{q,\ldots,\log^{\C{m}}N\cdot q\}$ and an estimate $x$ such that $x \in
[q'd-\frac{N}{\log^{\C{m}+1}N},q'd+\frac{N}{\log^{\C{m}+1}N}]~(\mod~N)$.
\end{lemma}
\begin{proof}\omittedproof{
Assume we are given a $q'$-matching $f$. We call routines $R_1$ and $R_2$ $\log^{3\C{m}+4}N$ times. If the number of
successful calls to one of the routines is less than $\log^{2\C{m}+3}N$, we fail. Otherwise, let $x\in [0,1]$ be the
average of the successful calls to $R_1$ and $y\in [0,1]$ be the average of the successful calls to $R_2$. According to
the Chernoff bound,
 $$Pr[|x-(\frac{1}{2}-\frac{1}{2}\cos(2\pi q' \frac{d}{N})) | > \frac{1}{{\C{e}}\log^{\C{m}+1}N} ] <
       2e^{-2 \log^{2\C{m}+3}N / ({\C{e}}^2\log^{2\C{m}+2}N)}$$
which is exponentially low in $\log N$ for any constant $\C{e}>0$. A similar bound holds for $y$. Hence, we can assume
that $x'=1-2x$ and $y'=2y-1$ are approximations of $\cos(2\pi q' \frac{d}{N})$ and of $\sin(2\pi q' \frac{d}{N})$
respectively up to an additive error of $\frac{2}{\C{e}\log^{\C{m}+1}N}$. According to Claim~\ref{sincos_estimate},
this translates to an estimate of $q' \frac{d}{N}~\mod~1$ with an additive error of $\frac{1}{\log^{\C{m}+1}N}$ for
$\C{e}$ large enough.

By repeating the above procedure with all the matchings that appear in Lemma~\ref{lemma_matching}, we are guaranteed to
find a good matching. According to Lemma~\ref{routine_a}, a call to routine $R_1$ or to routine $R_2$ with a good
matching succeeds with probability at least $\C{g}\frac{1}{\log^{\C{m}} N}$ for a certain $\C{g}>0$. The probability
that none of $\log^{\C{m}+1} N$ calls to the subroutine succeeds is $(1- \C{g}\frac{1}{\log^{\C{m}} N})^{\log^{\C{m}+1}
N}$ which is exponentially small. Thus, for one of the matchings, with probability exponentially close to $1$ we have
$\log^{2\C{m}+3}N$ successful calls to routines $R_1$ and $R_2$ and routine $R_3$ is successful.
}\end{proof}

We conclude the proof of Theorem~\ref{theorem_two_point} with a description of the algorithm for finding $d$. We begin
by using routine $R_3$ with the value $1$ to obtain an estimate $x_1$ and a value $\hat q\le \log^{\C{m}}N$ such that
$x_1 \in [d'-\frac{N}{\log^{\C{m}+1}N},d'+\frac{N}{\log^{\C{m}+1}N}]~(\mod~N)$ where $d'$ denotes $(d \hat q~\mod~N)$.
In the following we find $d'$ exactly by calling $R_3$ with multiples of $\hat q$. The algorithm works in stages. In
stage $i$ we have an estimate $x_i$ and a value $q_i$. The invariant we maintain is $x_i \in [q_i
d'-\frac{N}{\log^{\C{m}+1} N},~q_i d'+\frac{N}{\log^{\C{m}+1} N}]~(\mod~q_i N)$ . We begin with $x_1$ as above and
$q_1=1$. Assume that the invariant holds in stage $i$. We use routine $R_3$ with the value $2q_i \hat q$ to obtain an
estimate $x$ with a value $q' \in \{2q_i\hat q,4q_i\hat q,\ldots 2\log ^{\C{m}}N\cdot q_i\hat q\}$ such that $x\in
[q_{i+1} d'-\frac{N}{\log^{\C{m}+1}N},~q_{i+1} d'+\frac{N}{\log^{\C{m}+1}N}]~(\mod~N)$ where $q_{i+1}=q'/\hat q$.
Notice that our previous estimate $x_i$ satisfies $\frac{q_{i+1}}{q_i}x_i \in [q_{i+1} d'-\frac{2N}{\log N},q_{i+1}
d'+\frac{2N}{\log N}]~(\mod~q_{i+1}N)$. Since this range is much smaller than $N$, we can combine the estimate $x$ on
$(q_{i+1}d'~\mod~N)$ and the estimate $\frac{q_{i+1}}{q_i}x_i$ on $(q_{i+1}d'~\mod~q_{i+1}N)$ to obtain $x_{i+1}$ such
that $x_{i+1} \in [q_{i+1}d' - \frac{N}{\log^{\C{m}+1}N} ,~q_{i+1}d' + \frac{N}{\log^{\C{m}+1}N}] ~(\mod~q_{i+1}N)$.
The last stage is when $q_i \ge \frac{4N}{\log^{\C{m}+1} N}$. Then, $d'$ can be found by rounding $\frac{x_i}{q_i}$ to
the nearest integer. Given $d'$ there are at most $\hat q\le \log^{\C{m}}N$ possible values for $q$. Since this is only
a polynomial number of options we can output one randomly.

\section{Acknowledgements}
I would like to thank Dorit Aharonov, Noga Alon, Andris Ambainis, Irit Dinur, Sean Hallgren, Alexei Kitaev, Hartmut
Klauck, Ashwin Nayak, Cliff Smyth and Avi Wigderson for many helpful discussions and comments.

\bibliographystyle{plain}

\end{document}